\documentclass{aastex631}

\usepackage{amsmath}  % 必需宏包
\begin{document}

\title{A binary origin of ultra-long period radio pulsars}

\author[0000-0001-8356-2233]{Ying-Han Mao}
\affiliation{School of Astronomy and Space Science, Nanjing University, Nanjing 210023, P. R. China}
\affiliation{Key Laboratory of Modern Astronomy and Astrophysics (Nanjing University),
Ministry of Education, Nanjing 210023, P. R. China}

\author[0000-0002-0584-8145]{Xiang-Dong Li}\thanks{E-mail: lixd@nju.edu.cn}
\affiliation{School of Astronomy and Space Science, Nanjing University, Nanjing 210023, P. R. China}
\affiliation{Key Laboratory of Modern Astronomy and Astrophysics (Nanjing University),
Ministry of Education, Nanjing 210023, P. R. China}

\author[0000-0002-1934-6250]{Dong Lai}
\affiliation{Tsung-Dao Lee Institute, Shanghai Jiao Tong University, Shanghai 201210, P. R. China}
\affiliation{Center for Astrophysics and Planetary Sciences, Cornell University, Ithaca, NY 14853 USA}

\author[0000-0002-1398-2694]{Zhu-Ling Deng}
\affiliation{School of Astronomy and Space Science, Nanjing University, Nanjing 210023, P. R. China}
\affiliation{Key Laboratory of Modern Astronomy and Astrophysics (Nanjing University),
Ministry of Education, Nanjing 210023, P. R. China}

\author[0000-0001-5532-4465]
{Hao-Ran Yang}
\affiliation{School of Astronomy and Space Science, Nanjing University, Nanjing 210023, P. R. China}
\affiliation{Key Laboratory of Modern Astronomy and Astrophysics (Nanjing University),
Ministry of Education, Nanjing 210023, P. R. China}

\begin{abstract}
We propose a possible binary evolution model for the formation of ultra-long period pulsars (ULPPs). 
The model involves two key stages: first, a neutron star (NS) in wide binaries undergoes an effective spin-down phase through wind-fed accretion from its massive stellar companion; second, the supernova explosion of the companion leads to the disruption of the binary system, and produces two isolated compact stars. One of the them is the first-born, slowly rotating NSs, and our binary and spin evolution calculations show that the spin periods range from $\lesssim 0.1$ s to $\gtrsim 10^8$ s. This offers a possible formation channel for some of the long-period radio transients. We estimate that the formation rate of such systems in the Milky Way is approximately about $10^{-6}$ $\rm yr^{-1}$.
\end{abstract}


\section{Introduction} \label{sec:intro}
Radio pulsars are magnetized neutron stars (NSs) with spin periods typically ranging from milliseconds to several seconds. 
The initial spin periods primarily determined by the angular momenta of the pre-supernova cores of the progenitor stars, and depend on the physical processes of angular momentum transport during the massive stellar evolution \citep{Heger2005}. Current calculations suggest natal NS spin periods of 50-200 ms \citep{Ma2019}. Even with a negligible pre-SN core angular momentum, off-centered natal kick associated with asymmetric SN explosion may give the NS an initial spin with period of a few seconds \citep{Burrows2024}. 
However, the recent discovery of a growing population of ultra-long period radio transients (LPRTs) with pulse periods varying from a few $10$ to $\sim 10^4$ seconds \citep[e.g.,][]{Hurley-Walker2022,Hurley-Walker2023,Caleb2024} present an intriguing puzzle.  While these radio transients could be magnetic white dwarfs \citep{Katz2022,Loeb2022,Rea2024}, their emission properties suggest that they are likely magnetic NSs \citep[e.g.,][]{Men2025}.  Also, the central compact X-ray source at the center of the 2 kyr-old supernova remnant RCW 103 is known to have period of 6.7 hours \citep{DeLuca2006}.  The existence of these ultra-long-period pulsars (ULPPs) challenges the conventional pulsar formation theories.
These extremely long spin periods cannot be explained by traditional rotational energy loss through magnetic dipole radiation alone. 
For a NS born with initial period $P_0 = 10$ ms and magnetic field strength of $B \sim 10^{12}$ G ($10^{14}$ G), it would require $\sim 10^{13}$ yr ($10^{9}$ yr) to reach a period of $P_{\rm s} \sim 10^3$ s, which is clearly not feasible,  especially since of these sources are associated with SN remnants.
Consequently, the formation and evolution of ULPPs may involve extraordinary mechanisms.
Possible explanations include magnetars surrounded by a fallback disk formed from supernova ejecta \citep{DeLuca2006,Li2007,Xu2024,Yang2024}.

Most ULPPs are likely isolated objects. In comparison, long-period X-ray pulsars ($P_{\rm s} >$ a few $10$ s) have been commonly observed in wind-fed high-mass X-ray binaries (HMXBs) \citep[]{Neumann2023,Fortin2023}. For example, 4U 1954+31 \citep{Corbet2006} and 2S 0114+65 \citep{Finley1992} exhibit X-ray pulsations with periods $>10^3$ s.
Such long periods are thought to originate from the interaction of the NS magnetic fields with the stellar winds from the massive companion stars during the propeller and accretion processes \citep{Illarionov1975,Shakura2012}. Angular momentum transfer from the NS to the wind material can result in efficient spin-down over time,  especially for unsteady wind accretion \citep{Mao2024}.

As time progresses, the companion star in such HMXBs will evolve toward the end of its life and undergo a supernova explosion.
Sudden mass loss and natal kick induced by the asymmetry of the supernova explosion may lead to the disruption of the binary system. 
For the systems where the NSs have already experienced significant spin-down before the supernova explosion, an isolated NS with long spin period is produced, which can potentially appear as ULPPs.
This may offer an alternative formation channel for ULPPs.

In this work, we quantify the binary origin hypothesis for ULPPs. 
In Section \ref{sec:model}, we describe the formation process of ULPPs in binary systems and the spin evolution model for NSs.  
Section \ref{sec:result} presents the distribution of the final spin periods and the birth rate derived from binary population synthesis (BPS) and Monte Carlo (MC) simulations. 
We conclude in Section \ref{sec:conclusion}.

\section{Model} \label{sec:model}
\subsection{Binary Formation Channel for ULPPs} \label{subsec:2.1}
\begin{figure}[ht!]
\centering
\includegraphics[height=18cm]{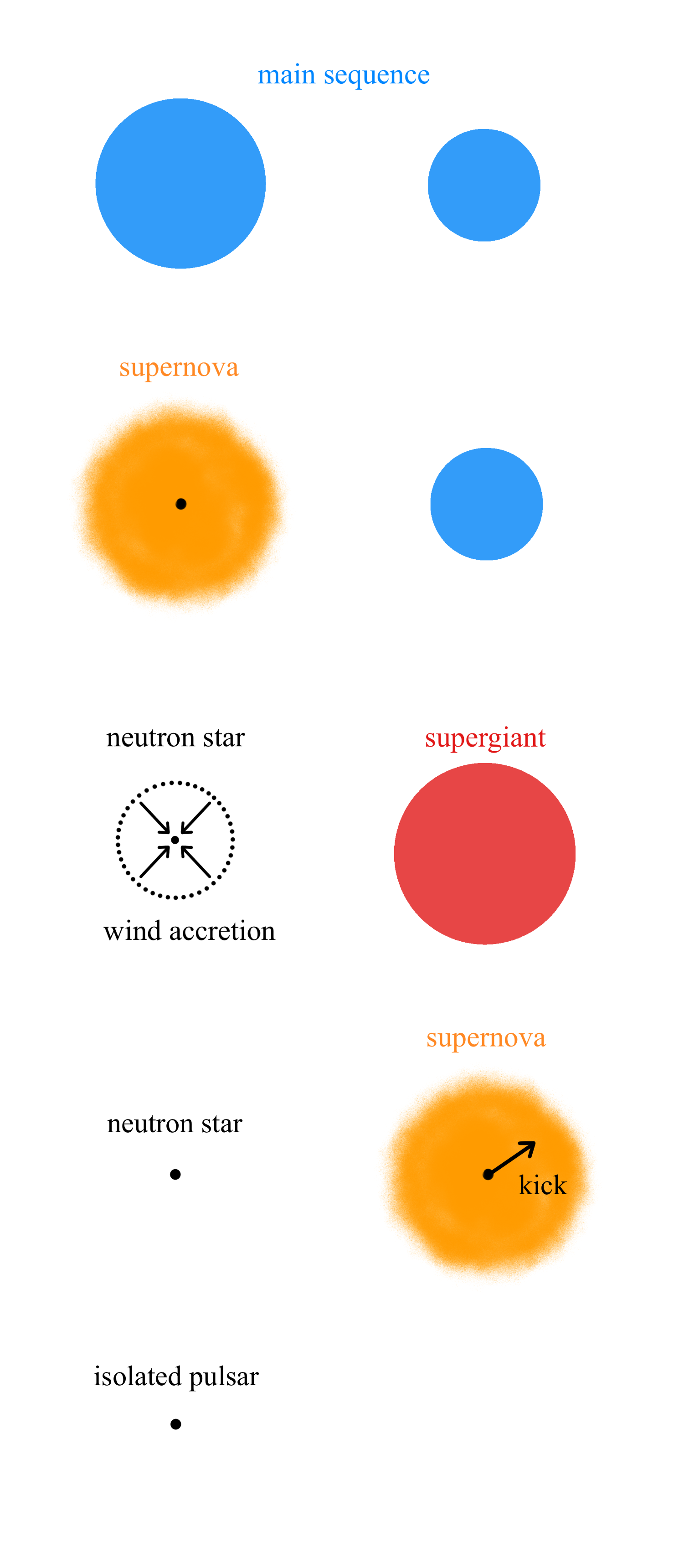}  
\caption{The evolutionary path of the formation of ULPPs in a binary system. 
The initial phase is a main-sequence binary system, where the more massive primary star on the left undergoes a supernova explosion, forming a NS (Step 1 - Step 2). 
This NS then accretes material from the wind of the companion star through a wind-fed process, resulting in angular momentum loss and spin-down (Step 3). 
When the companion star on the right eventually reaches the supernova phase, the violent mass ejection disrupts the binary system (Step 4).
At this point, the first-born NS is left behind, eventually becoming an isolated NS (Step 5).
Its slow rotation period makes it a potential candidate for ULPPs.
\label{fig:figure1}}
\end{figure}

Figure \ref{fig:figure1}  depicts the binary evolutionary path that leads to the formation of ULPPs. 
The process begins with a binary system consisting of two massive main-sequence (MS) stars (Step 1). 
As the primary star evolves, it reaches the end of its life and undergoes a supernova explosion producing an NS (Step 2). 
In the case that the binary system remains gravitationally bound, the companion evolves into a supergiant with strong winds, and the NS spins down when accreting from the wind material (Step 3). In systems with sufficiently wide orbits, Roche-lobe overflow does not occur, and the NS maintains its slow spin until the companion star undergoes the supernova explosion.
The second formed NS or black hole is imparted with a natal kick, which may disrupt the binary (Step 4).
The first-born NS could appear as a ULPP (Step 5).

In the following, we employ the NS spin evolution model in wind-fed HMXBs, along with BPS simulation, to calculate the spin evolution during each stage of this process and estimate the birth rate of ULPPs through this channel.

\subsection{Spin Evolution Model of NSs in Wind-Fed Systems} \label{subsec:2.2}
At Step 3 shown in Figure \ref{fig:figure1}, the NS is embedded within the stellar wind of the companion star and accretes material from it. 
This process is described as wind-fed accretion.
The accretion process of the NS is determined by both the gravitational force and the magnetic fields, depending on three characteristic radii -- the magnetospheric radius $R_{\rm m}$, the light cylinder radius $R_{\rm lc}$, and the corotation radius $R_{\rm co}$.
Define the NS mass as $M_{\rm NS}$, radius as $R_{\rm NS}$, magnetic field as $B$, and magnetic moment $\mu = B R_{\rm NS}^{3}$, with the accretion rate denoted as $\dot{M}$. 
The magnetosphere radius $R_{\rm m}$ is given by \citep{Lamb1973,Fabian1975}
\begin{equation}
    R_{\rm m}=\left(\frac{\mu^2}{2\dot{M}\sqrt{2GM_{\rm NS}} }\right)^{2/7},
\end{equation} 
where $G$ is the gravitational constant.
This radius is determined by balancing the ram pressure of the accretion flow with the magnetic pressure, and sets the inner radius of the accretion flow.

The light cylinder radius  $R_{\rm lc}$ is defined as
\begin{equation}
    R_{\rm lc}=\frac{cP_{\rm s}}{2 \pi},
\end{equation}      
where $c$ is the speed of light.
At the location of the light cylinder radius, when the material rotates with the angular velocity of the NS, its tangential velocity reaches the speed of light. 
This radius defines the outer boundary of the magnetosphere. 
Beyond this critical radius, the magnetic field lines transition from a closed to an open configuration.

The corotation radius $R_{\rm co}$ is defined as the radial distance at which the spin angular velocity of the NS equals the Keplerian angular velocity,
\begin{equation}
    R_{\rm co}=\left(\frac{GM_{\rm NS}P_{\rm s}^{2}}{4 \pi^{2}}\right)^{1/3}.
\end{equation} 
At the corotation radius, the gravitational force acting on the corotating plasma is balanced by the centrifugal force.
Beyond this radius, the centrifugal force dominates.

The accretion phase of the NS is determined by the relationships between these three radii.
There are considerable uncertainties in the torque acting on the NS in different accretion phases, particularly when the NS reaches long spin period.  In the following, as an illustration, we largely follow work described in \cite{Lipunov1992} and \cite{Shakura2012} to calculate the spin-down torques on the NS -- the same torques have been used to explain the long-period pulsars in HMXBs.

A newborn NS likely has a rapid spin with $R_{\rm m}>R_{\rm lc}$. 
The gravitationally captured wind matter cannot penetrate into the magnetosphere and interact with the NS.
During this period, the NS is in the $ejector$ phase ($phase ~a$), emitting ``pulsar wind" and spinning down by magnetic dipole radiation.  The torque exerted on the NS is given by 
\begin{equation}
    N_{\rm a}\simeq -\frac{16\pi^{3}\mu^{2}}{3 c^{3}P_{\rm s}^{3}}.
	%\label{eq:quadratic}
\end{equation}   

As the NS spins down, $R_{\rm lc}$ gradually increases and becomes greater than $R_{\rm m}$, allowing matter to enter the magnetosphere. 
If $R_{\rm co}<R_{\rm m}<R_{\rm lc}$, the centrifugal force on the corotating matter surpasses gravity,  accretion onto the NS is inhibited, and the matter is either expelled or stalled at the boundary \citep{Illarionov1975}.
This stage is called the $propeller$ phase ($phase~b$). The torque generated during this process is given by \citep{Shakura1975}
\begin{equation}
    N_{\rm b} \simeq -\dot{M}R_{\rm m}^{\rm 2}\left(\frac{2\pi}{P_{\rm s}}\right).
\end{equation}  

Mass ejection causes the NS to rapidly lose angular momentum, resulting in a significant increase in $P_{\rm s}$.
Once $P_{\rm s}$ grows sufficiently to reach the condition $R_{\rm m}\leq R_{\rm co}$, matter can accrete onto the NS surface, and the star enters the $accretor$ phase. 
This phase is divided into two different cases: Bondi-Hoyle accretion ($phase~c$) and subsonic settling accretion ($phase~d$).
These two cases are separated by the critical X-ray luminosity $L_{\rm crit} = 4\times10^{36}\mu_{30}^{1/4}~{\rm erg}~{\rm  s^{-1}}$ \citep{Postnov2011}. Bondi-Hoyle accretion applies when $L_{\rm X}>L_{\rm crit}$. In this case the accreting matter is efficiently cooled through Compton scattering and enters the magnetosphere supersonically. The accretion rate of the wind-fed system can be written as \citep{Bondi1944}
\begin{equation}
    \dot{M}=\pi R_{\rm G}^{2} \rho v_{\rm rel}=\frac{\pi R_{\rm G}^{2}\dot{M}_{\rm w}v_{\rm rel}}{4\pi a^{2}v_{\rm w}},
\end{equation}
where $\rho$ is the stellar wind density, $v_{\rm rel}=(v_{\rm w}^{2}+v_{\rm orb}^{2})^{1/2}$ is the relative velocity, $a$ is the binary semi-major axis, $\dot{M}_{\rm w}>0$ is the wind mass loss rate, $R_{\rm G}={2GM_{\rm NS}}/{v_{\rm rel}^{2}}$ is the gravitational capture radius of the NS.
The stellar wind velocity of the companion star is \citep{Castor1975}
\begin{equation}
    v_{\rm w}=v_{\rm \infty}(1-\frac{R_{2}}{a})^{\beta}=\alpha v_{\rm esc}(1-{R_{2}\over{a}})^{\beta},
\end{equation}
where $v_{\rm \infty}$ is the terminal wind velocity, $v_{\rm esc}=\sqrt{2GM_{2}/R_{2}}$ is the escape velocity from the companion star. 
The coefficient $\alpha$ and the power law index $\beta$ are set to be 1 and 0.8, respectively \citep{Waters1989}. 

The accreting X-ray luminosity is given by
\begin{equation}
    L_{\rm X}=\eta\frac{GM_{\rm NS}\dot{M}}{R_{\rm NS}}.
\end{equation} 
where $\eta=0.1$ is the conversion efficiency, representing the fraction of gravitational energy of the accreted matter that is converted into X-ray luminosity.

In the other case of subsonic settling accretion, Compton cooling is inefficient and the radial velocity of the accreting matter is subsonic, forming a hot shell outside the NS. Mass accretion occurs through the Rayleigh–Taylor instability, so
the actual accretion rate $\dot{M}_{\rm acc}$ is substantially smaller than $\dot{M}$, and expressed as \citep{Postnov2011}.\footnote{All subscript numbers in this paper represent powers of 10, e.g.  $\dot{M}=\dot{M}_{16}\times10^{16}~ {\rm g~s^{-1}}$.}
\begin{equation}
{\dot{M}_{\rm acc}\over{\dot{M}}}=0.3\dot{M}_{16}^{4/11}\mu_{30}^{-1/11}.
\end{equation}

In summary, the torque acting on the NS in the $accretor$ phase is determined by the interaction between the magnetosphere and the accreting matter, together with the angular momentum transfer from the infalling matter.
Considering both effects, the torque acting on the NS can be expressed as \citep{Popov1999,Popov2012},
\begin{equation}
     N_{\rm c}=f \zeta \dot{M} R_{\rm G}^{\rm 2} \left(\frac{2\pi }{P_{\rm orb}}\right)-\frac{\mu^{2}}{3R_{\rm co}^{3}}~~~~~~~~{\rm if}\ L_{\rm X}>L_{\rm crit},
\end{equation} 
\begin{equation}
    N_{\rm d} =f A\dot{M}_{\rm acc,16}^{7/11}-B\dot{M}_{\rm acc,16}^{3/11}
    ~~~~~~~~~~~{\rm if}\ L_{\rm X}<L_{\rm crit}.
\end{equation} 
where 
\begin{equation}
A=4.60\times10^{31}K_{1}\mu^{1/11}_{30}v_{8}^{-4}\left(P_{\rm orb}/{10~\rm day}\right)^{-1},  
\end{equation} 
\begin{equation}
B=5.49\times10^{32}K_{1}\mu^{13/11}_{30}\left(P_{\rm s}/{100~\rm s}\right)^{-1},
\end{equation} 
and $\zeta=0.25$ and $K_{1}=40$ are the dimensionless coefficients; 
$f$ is a parameter used to average the alternation of the direction of the wind matter's angular momentum with typical value of 0.1 \citep{Mao2024}.

The time derivative of the spin period is
\begin{equation}
   \dot{P}=- \frac{P_{\rm s}^{2}}{2\pi I}N,
\end{equation}  
where $I = 10^{45} \rm~g~cm^{2}$ is the moment of inertia of the NS. Since the initial period $P_0$ has minor impact on the final result, we set $P_0 = 0.2$ s in all calculations. Combining stellar and binary evolution, we can follow the spin evolution of NSs during their lifetimes, based on the above torque equations.

\section{Results} \label{sec:result}

\subsection{Spin Period Distribution in the $M_2-P_{\rm orb}$ Parameter Space } \label{subsec:tables}

We use the stellar evolutionary code Modules for Experiments in Stellar Astrophysics ({\tt MESA}) \citep{Paxton2011,Paxton2013,Paxton2015,Paxton2019} to simulate the evolution of the companion star from zero-age main-sequence to supernova explosion.
The metallicity is fixed to the solar value ($Z=0.02$) \footnote{\textcolor{black}{The data is available on Zenodo under an open-source license:\dataset[doi:10.5281/zenodo.15621133]{https://doi.org/10.5281/zenodo.15621133}.}
}.
The stellar wind mass-loss rate is determined using the empirical formulae from \cite{Vink2001} for stars with $T_{\rm eff}>1.1\times10^4$ K and from \cite{deJager1988} for those with $T_{\rm eff}<1.1\times10^4$ K. 
The latter is also applied to stars with a central hydrogen abundance below 0.01 and a central helium mass fraction below $10^{-4}$. 
During the HMXB stage, we employ the Roche lobe radius $R_{\rm L}$ of the companion star \citep[]{Eggleton1983}
\begin{equation}
   {R_{\rm L}}=\frac{0.49q^{2/3}a}{0.6q^{2/3}+{\rm ln}(1+q^{1/3})},
	%\label{eq:quadratic}
\end{equation}
where $q=M_{2}/M_{\rm NS}$ is the ratio of the companion and NS masses, to determine the accretion modes. 
When $R_2\geq R_{\rm L}$, Roche-lobe overflow occurs, resulting in the formation of an accretion disk around the NS. Since disk accretion can rapidly spin up the NS, we terminate the calculation at this point and discard the corresponding binary.

\begin{figure}[ht!]
\centering
\includegraphics[width=\textwidth]{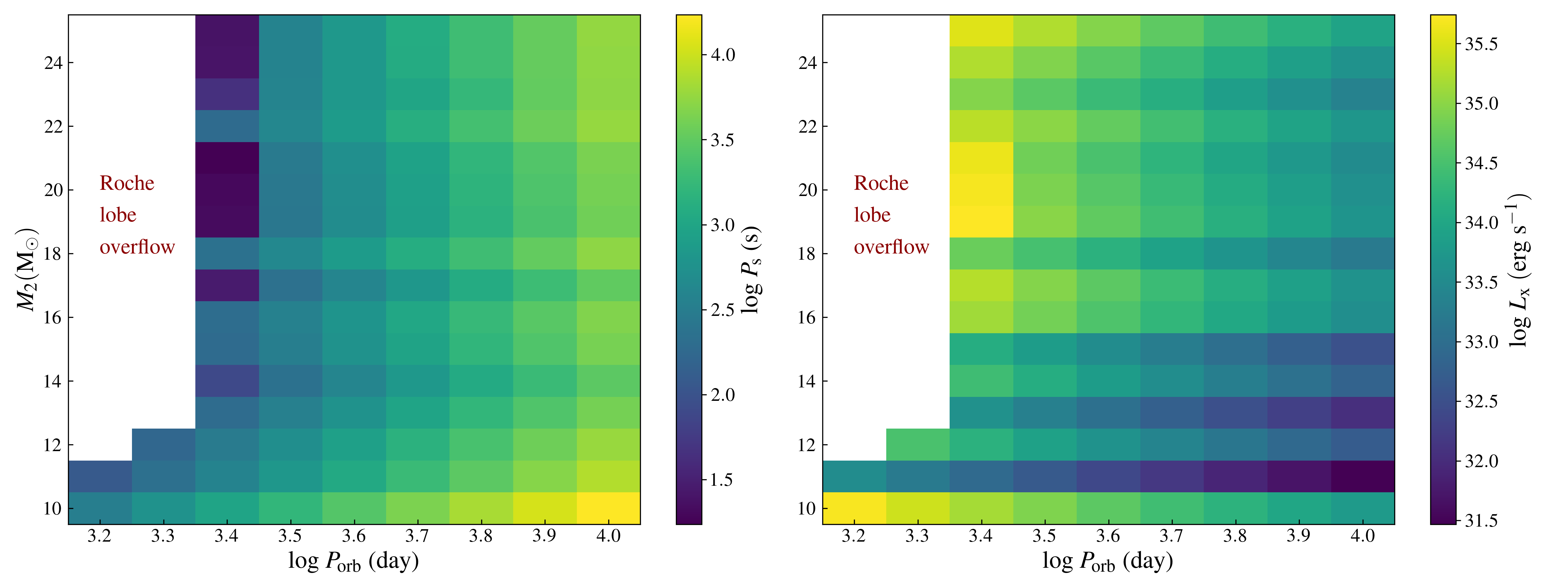}  
\caption{The distribution of final spin period ($P_{\rm s}$) and X-ray luminosity ($L_{\rm X}$) in the parameter space of companion (ZAMS) star mass and orbital period.
The magnitudes of $P_{\rm s}$ and $L_{\rm X}$ are displayed with different colors. The white area indicates systems that experience Roche lobe overflow. 
\label{fig:figure2}}
\end{figure}

For NSs whose companions do not fill the Roche lobes during the whole accretion stage, we combine the mass-loss rate $\dot{M}_{\rm w}$ with the torque model to follow the NS spin evolution until the supernova explosion of the companion. The results are presented in Figure \ref{fig:figure2}.
The horizontal and vertical axes represent the initial $P_{\rm orb}$ in logarithm and $M_2$, respectively. 
The orbital period ranges from 1,000 to 10,000 days with a logarithmic step of 0.2, and the initial companion mass ranges from 10 to 25 $\rm M_{\odot}$ with a step of 1 $\rm M_{\odot}$.
For each parameter set, the spin period $P_{\rm s}$ of the NS and its corresponding X-ray luminosity $L_{\rm X}$ at the time of the second supernova are represented by the color of each grid cell in Figure \ref{fig:figure2}. 
The white regions indicate systems that will undergo Roche-lobe overflow.
The range of $P_{\rm s}$ spans from a few ten to more than $10^4$ seconds, which well covers the observed pulse periods of ULPPs.
The right panel of Figure \ref{fig:figure2} demonstrates the expected correlation: higher companion masses and shorter orbital periods result in higher accretion rates. 
NSs in systems with lower mass companions and wider orbits generally evolve into longer spin periods due to weaker accretion.

\begin{figure}[ht!]
\centering
\includegraphics[height=15cm]{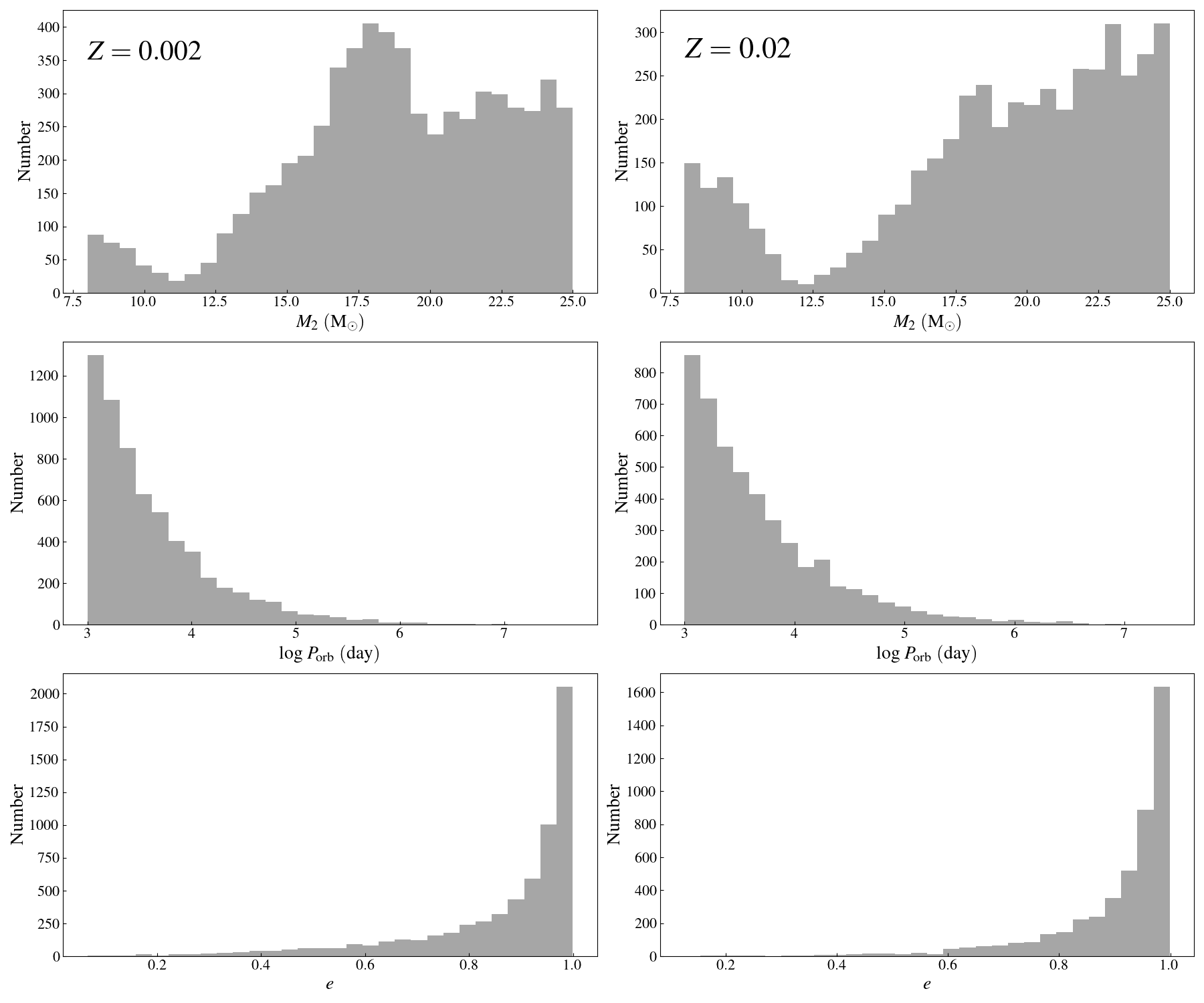}  
\caption{The number distribution of the companion star mass ($M_2$), orbital period ($P_{\rm orb}$), and eccentricity ($e$) of the binary systems selected from the BPS simulation results, after the primary star has exploded and become a NS. \textcolor{black}{The left and right panels correspond to the results for $Z = 0.002$ and $Z = 0.02$, respectively.} This distribution corresponds to the parameters of the initial binary systems in Step 2 of Figure \ref{fig:figure1} and reflects the statistical weight of the initial systems retained after the evolutionary simulation.  
The noticeable dip in the $M_2$ distribution around 12 $\rm M_\odot$ is likely caused by the mass exchange within binary systems.
Primaries with mass $>$ 10 $\rm M_\odot$ are more likely to initiate mass transfer.  During this stage, the companion stars typically gain a significant amount of mass, shifting them into the $M_2>15~\rm M_\odot$ range.}
\label{fig:figure3}
\end{figure}

\subsection{Monte Carlo simulations of the Final NS Spin Periods}

We further combine binary population synthesis and Monte Carlo simulations to determine the birthrate of ULPPs.
We first use the BPS code, originally developed by \cite{Hurley2000,Hurley2002} to simulate the evolution of a large number of main-sequence binary systems in the Galaxy.  We adopt a star formation rate of 3 $\rm M_{\odot} ~yr^{-1}$ for the Milky Way, or 6.6 star $\rm yr^{-1}$ for stars in the mass range of 0.08 – 100 $\rm M_{\odot}$ according to the initial mass function (IMF) \citep{Kroupa1993,Chomiuk2011}. 
The initial parameters for the BPS model are set as follows. The metallicity is set to the solar value $Z=0.02$ (\textcolor{black}{the influence of metallicity will be discussed below}). 
The primary star mass $M_1$ follows the \citet{Kroupa1993} IMF within a mass range of 8 to 60 $\rm M_{\odot}$.
The mass ratio $q=M_2/M_1$ is uniformly distributed between 0 and 1 \citep{Kobulnicky2007}.
The initial orbital separation $a$ is drawn from a logarithmic uniform distribution ranging from 3 to $10^4~\rm R_{\odot}$ \citep{Abt1983}.
All binary systems are assumed to reside initially in circular orbits.

We terminate the BPS simulation when the primary star undergoes a supernova explosion. 
Supernova explosions occur in two ways: core-collapse supernovae (CCSNe) and electron-capture supernovae (ECSNe). The latter typically occur in stars with initial masses ranging from approximately 8 to 10 $\rm M_\odot$. 
Here, we assume that a star will explode in an ECSN if the mass of its helium core is in the range of 1.83 to 2.25 $\rm M_\odot$ \citep{Shao2018}.
We also assume that the velocity distribution of newborn NSs follows a Maxwellian distribution, with a velocity dispersion of 265 km $\rm s^{-1}$ for CCSNe \citep{Hobbs2005} and 30 km $\rm s^{-1}$ for ECSNe \citep{Podsiadlowski2004,Verbunt2017,Deng2024}. \textcolor{black}{Figure 3 (right panel) presents the distribution of the companion star mass ($M_2$), orbital period ($P_{\rm orb}$), and eccentricity ($e$) in the binary systems immediately after the primary star becomes a NS.}
 
According to these probability distributions, we generate initial NS + MS binary systems (with a step of 0.2 $\rm M_\odot$ for $M_2$), and derive the evolution of stellar wind mass loss and $R_2$  from {\tt MESA} calculations . Based on these values, we calculate the spin evolution of the NS in each system. We assume that the initial NS magnetic fields follow a logarithmic normal distribution with $\mu_{{\rm log}~(B/{\rm G})}=12.65$ and $\sigma_{{\rm log}~ (B/{\rm G})}=0.55$ \citep{Faucher2006}.
Note that an eccentricity is induced after the supernova and this should have an impact on the accretion process.  We replace the binary separation $a$ with the average distance $r$ given by
\begin{equation}
   r=a\sqrt{1-e^2},
	\label{eq:quadratic}
\end{equation}
to estimate the average NS accretion rate.

\begin{figure}[ht!]
\centering
\includegraphics[width=\textwidth]{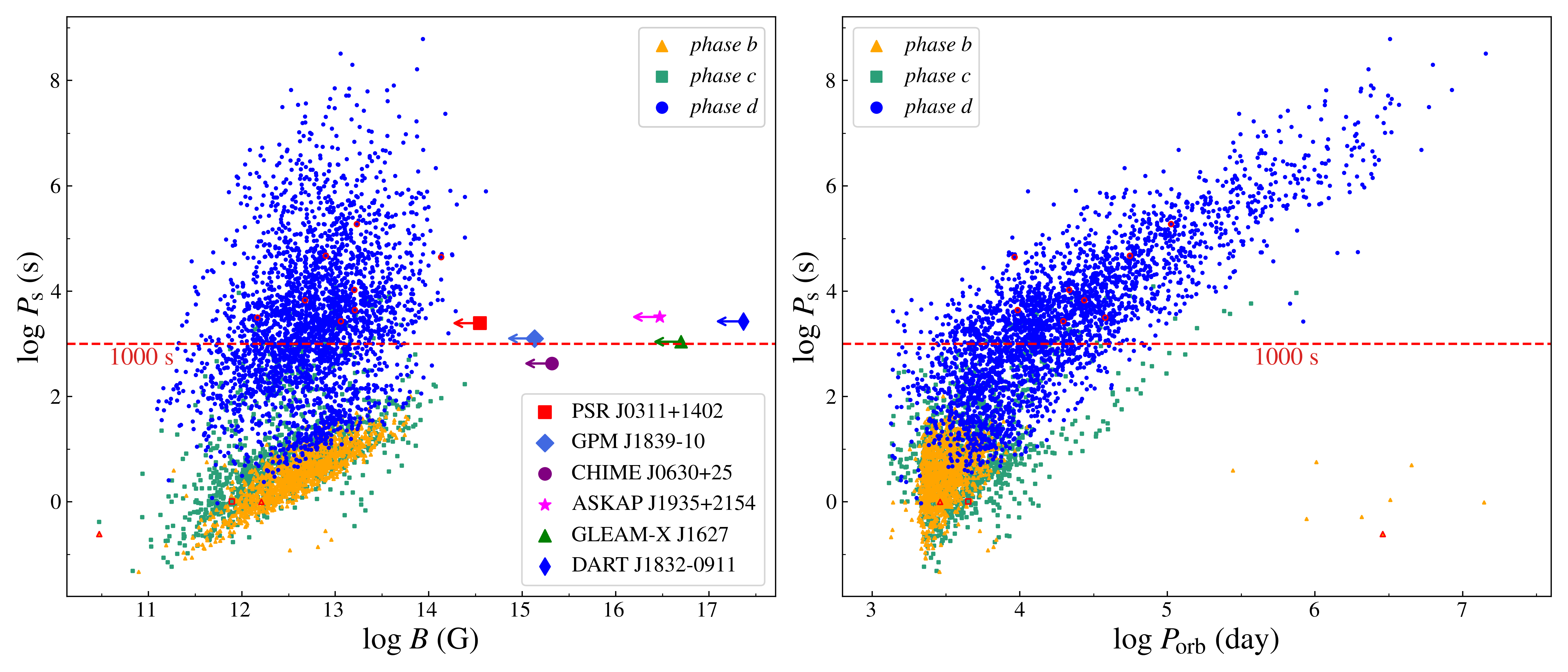}  
\caption{The distribution of the final NS spin period ($P_{\rm s}$) as a function of magnetic field strength ($B$) and orbital period ($P_{\rm orb}$).
The orange triangles, green squares, and blue dots denote systems in $phases~b$, $c$, and $d$, respectively. 
The small red circles are the systems that remain gravitationally bound after the second supernova explosion. 
Several known ULPPs are displayed in the left panel, with their magnetic fields representing the upper limits of the dipole field.
The red dashed line marks $P_{\rm s}=1000$ s.
\label{fig:figure4}}
\end{figure}

Our calculation stops when the secondary star evolves to a supernova. With the kick module of the BPS model, we calculate the probability of binary disruption due to the second supernova and find that it exceeds 99$\%$. This means that most NSs become isolated after the second supernova explosion (Step 4 - Step 5 in Figure \ref{fig:figure1}), and only 
a very small number of systems can survive and become double NSs. The latter are highlighted with the red circles shown in Figure \ref{fig:figure4}.
The left and right panels of Figure \ref{fig:figure4} depict the distribution of the first born NSs in the $P_{\rm s}-B$ and $P_{\rm s}-P_{\rm orb}$ planes, respectively. 
In these plots, the orange triangles, green squares, and blue dots represent NSs in $phases~b$, $c$, and $d$ at the moment of the second supernova, respectively.
In the left panel, symbols with different colors represent the currently known ULPPs, with the derived upper limits of their dipole field strengths from the observational data \citep{Hurley-Walker2022,Hurley-Walker2023,Dong2024,Caleb2024,Li2024,Wang2025}.
NSs in $phases~b$ and $c$ are concentrated in the lower region with smaller $P_{\rm s}$.
The narrow, strip-like distribution of $phase~b$ NSs arises from a reduction in $\dot{M}_2$ in the oxygen and silicon burning phase. During this stage, the outer layers have been largely stripped away, leading to a decrease in the stellar wind mass-loss rate. As a result, the magnetospheric radius expands, causing the NS to enter the 
$propeller$ stage. 
$Phase~c$ NSs result from systems with shorter orbital periods, thus having higher accretion rates ($L_{\rm X}>L_{\rm crit}$) with Bondi accretion.
This is also evident in the right panel of Figure \ref{fig:figure4} that both $phases~b$ and $c$ NSs are concentrated in the lower left corner.
NSs in $phase~d$ have experienced quasi-spherical subsonic accretion in HMXBs. In our calculations approximately $20\%$ NSs have reached $P_{\rm s}>1000$ s. We see that in binary systems with long orbital periods, the formation of slowly rotating NSs is a natural outcome.

%\subsection{The Birthrate of ULPPs from Binary Systems}

Based on the survival rate calculations in the previous steps, we estimate the total formation rate of NSs originating from binary systems to be $8\times 10^{-3}$ yr$^{-1}$, and the birthrate of potential ULPPs with spin periods exceeding 1000 s to be $1.4\times 10^{-6}$ yr$^{-1}$. 

\begin{figure}[ht!]
\centering
\includegraphics[width=\textwidth]{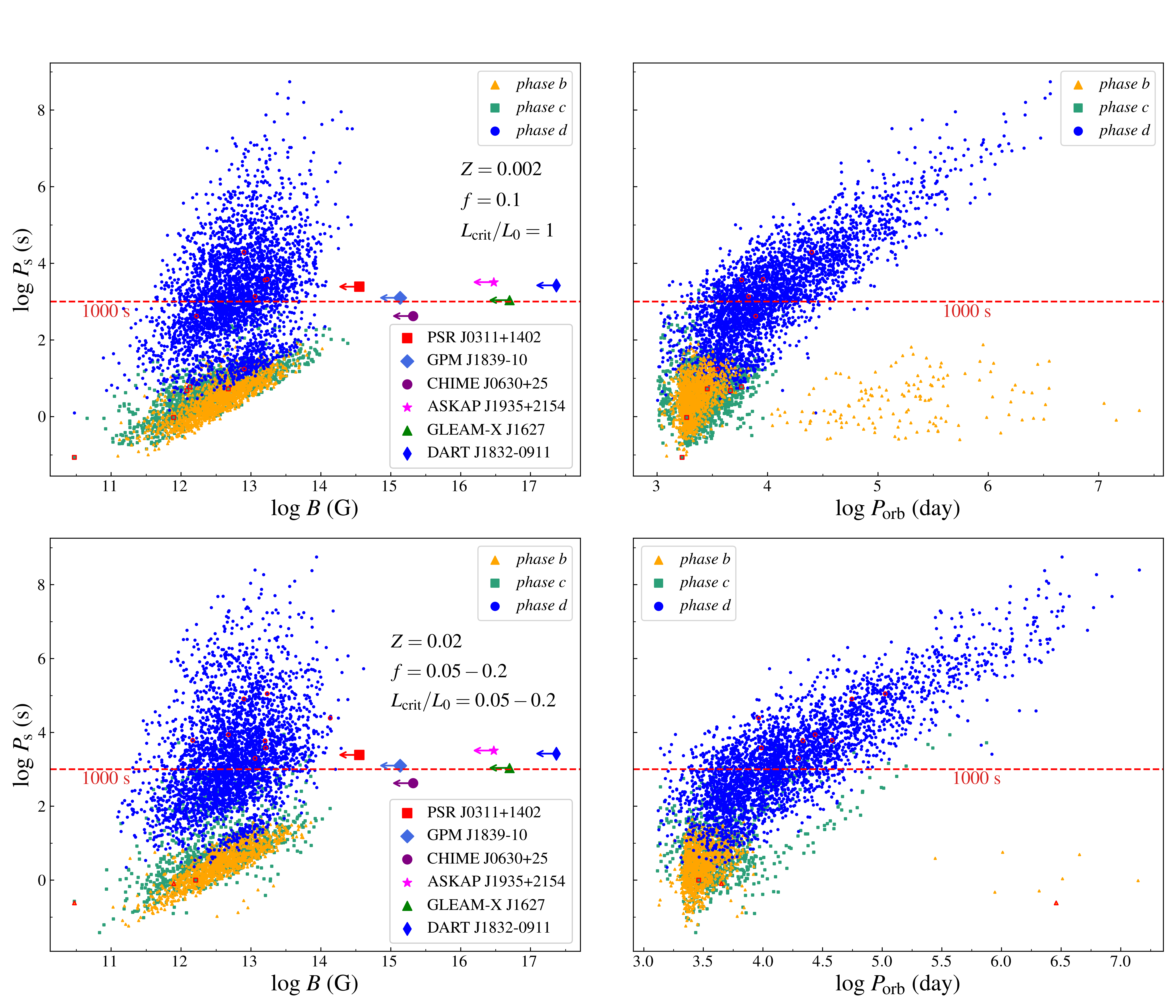}  
\caption{\textcolor{black}{Monte Carlo simulation results for sub-solar metallicity ($Z = 0.002$), fixed values of $f$ and $L_{\rm crit}/L_0$ (top panels) and for solar metallicity ($Z = 0.02$) and randomized values of $f$ and $L_{\rm crit}/L_0$ (bottom panels). All other settings are the same as in Figure \ref{fig:figure4}.}
\label{fig:figure5}}
\end{figure}

\textcolor{black}{To investigate how various physical factors influence the birthrate of binary-origin ULPPs, we perform a series of MC simulations with sub-solar metallicity $Z = 0.002$ and a range of dimensionless coefficients in the torque formula. Metallicity plays a crucial role in shaping a star’s lifetime, structure, and fate by altering opacity, nuclear reaction rates, mass loss, and remnant formation \citep[e.g.,][]{2001A&A...373..555M,Vink2001,Mokiem2007}.
Lower metallicity may modify the distribution of the BPS outcomes for NS binary formation and reduce the stellar wind mass-loss rate ($\dot{M}_{\rm w}$) 
during the subsequent spin evolution. Regarding the first effect, Figure~\ref{fig:figure3} shows that sub-solar metallicity has minor impact on the distributions of $M_2$, $P_{\rm orb}$, and $e$, with the overall shapes and peak positions remaining nearly unchanged compared to the solar-metallicity case. As for the second effect, while sub-solar metallicity stars experience lower radiative mass-loss due to reduced line opacity, this does not significantly affect the spin evolution of the NS. The final spin period in $phase~d$ depends relatively weakly on the wind mass-loss rate, approximately following $P_{\rm s} \propto \dot{M}_{\rm w,16}^{-0.5}$.
This weak dependence is consistent with the simulation results shown in the top panels of Figure~\ref{fig:figure5}, where the overall spin period distributions under sub-solar and solar metallicities are broadly similar. The impact of reduced mass-loss rates is most evident in wide-orbit systems, where an increased number of sources enter the $propeller$ phase (as shown in the top-right panel of Figure~\ref{fig:figure5}). This is because a lower $\dot{M}$ increases the critical spin period between the $propeller$ and $accretor$ phases ($P_{\rm crit} \propto \dot{M}^{-3/7}$), pushing more systems into the $propeller$ regime and delaying their transition to the accretion phase.
Based on the MC simulations at sub-solar metallicity, we estimate the birthrate of ULPPs with $P_{\rm s}>1000$~s to be approximately $1.9 \times 10^{-6}~{\rm yr}^{-1}$.
}

\textcolor{black}{In addition to metallicity, the dimensionless coefficients in the accretion torque formula may also influence the MC results. We simulate the spin period distributions with random values of $f$ ranging from 0.05 to 0.2 and $L_{\rm crit}/L_0$ (where $L_0=4\times10^{36}\mu_{30}^{1/4}~{\rm erg}~{\rm  s^{-1}}$) from 0.5 to 2 \citep{Xu2019,Postnov2011,Shakura2012}, as shown in the bottom panels of Figure~\ref{fig:figure5}.
Theoretically, a smaller $f$ leads to a longer spin period during the $accretor$ phase, while a lower $L_{\rm crit}$ tends to shift more systems with low accretion rates into $phase~c$; the opposite holds for larger $f$ and higher $L_{\rm crit}$. In the simulations with randomized $f$ and $L_{\rm crit}$, these effects largely offset each other, resulting in the spin period distributions that are overall comparable to those obtained with fixed parameters.
Interestingly, the standard deviation of the spin period distribution in the randomized case is approximately 90\% of that in the fixed-parameter case, indicating that the results with randomized parameters are actually more concentrated rather than more dispersed. This occurs because, in the fixed-parameter model, a small fraction of systems evolve to extremely long spin periods, increasing the overall dispersion. In contrast, randomized parameters suppress such extreme values, producing a smoother and more concentrated distribution. However, since systems with $P_{\rm s} > 1000$ s do not lie in the extreme tail of the distribution, the estimated birthrate of ULPPs remains largely unaffected, staying on the order of $\sim 10^{-6}~{\rm yr}^{-1}$.
}

%Among them, the formation rate of high-mass stars (8 – 60 $\rm M_{\odot}$) is 0.015149 star $\rm yr^{-1}$, accounting for approximately 0.23\% of the total star formation rate. 
%As a result, the formation rate of BPS initial input systems with primary star masses between 8 and 60 $\rm M_{\odot}$ corresponds to 0.015149 star $\rm yr^{-1}$.
%After the primary stars undergo supernova explosions, approximately 35\% collapse into NSs, but only about 3\% of them remain in binary systems.
%A total of 4668 systems are selected based on the criteria $8~{\rm M_\odot}<M_2<25~\rm M_\odot$, $P_{\rm orb}<1000$ day, and $0<e<1$, resulting in a survival probability $P_1= 4.67 \times 10^{-4}$ from Step 1 to Step 2 in Figure \ref{fig:figure1}.
%Following the first-stage selection, the spin periods of NSs in these surviving systems were further simulated when the companion star underwent a supernova explosion. 
%After the second-stage selection, the number of NSs with a spin period greater than 1000 s is calculated to be 1944 out of 10,000 systems, giving a corresponding probability $P_2$ of 0.1944.
%By combining the probabilities from both stages of selection, the annual formation rate of long-period NSs is ultimately calculated to be $1.41\times 10^{-6}$ star yr$^{-1}$. 
%To estimate the total number of such NSs in the Milky Way, assuming a lifetime of $10^9$ years, the number of ULPPs formed through this channel is $\sim$ 1400.

\begin{figure}[ht!]
\centering
\includegraphics[width=\textwidth]{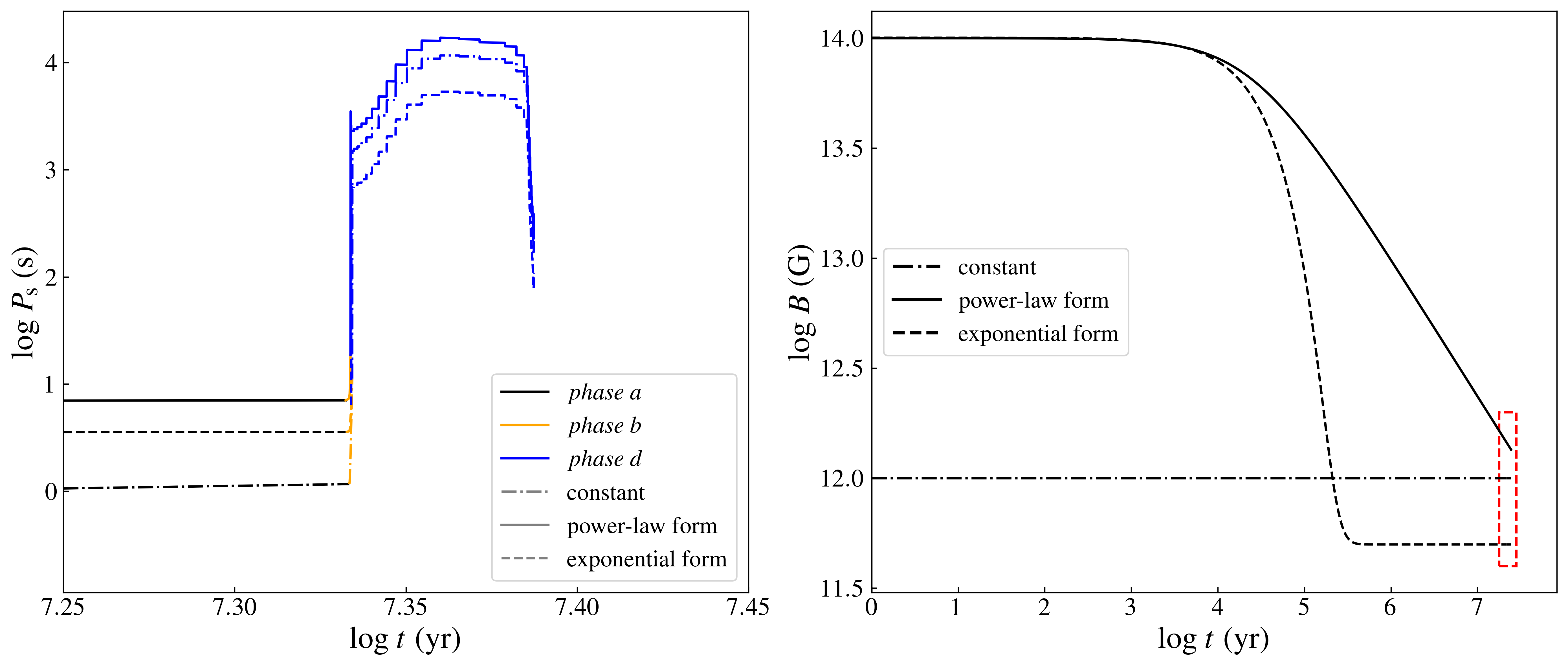}  
\caption{Evolutionary tracks of spin period ($P_{\rm s}$) and magnetic field ($B$) for a system with $M_2 = 10~{\rm M}_{\odot}$ and $P_{\rm orb} = 5000$ day.  The solid, dashed, and dotted lines represent three different magnetic field evolution scenarios: exponential decay, power-law decay, and constant field, respectively. In the left panel, the black, orange, and blue lines correspond to $phases$ $a$, $b$, and $d$, respectively. In the right panel, the red dashed box highlights the range of magnetic field strengths corresponding to the time interval shown in the left panel.
\label{fig:figure6}}
\end{figure}

Studies on the radio emission mechanisms have generally proposed that ULPPs are essentially magnetars with magnetic fields of $10^{14} - 10^{15}$ G \citep{Chen1993,Suvorov2023}.  However, considering magnetic field decay due to Ohmic dissipation and magnetic field burying resulting from material deposition during the accretion phase, it is difficult for binary origin ULPPs to maintain a strong field strength \citep{Cumming2001}.  For a newborn NS with the \textcolor{black}{initial magnetic field $10^{14}$ G}, its surface magnetic field would decay to the level of typical NSs ($10^{12}-10^{13}$ G) after $\sim 10^6$ years.  And the final spin period of the ULPPs is found to be largely depend on the terminal magnetic field strength, insensitive to the ways of field evolution. Figure \ref{fig:figure6} shows the magnetic field and spin period evolutionary tracks under three different magnetic field evolution scenarios \citep{Colpi2000}:
\begin{equation}
B(t) = \left\{
    \begin{aligned}
        & B_{0}\left(1+\gamma t/\tau \right)^{-1/\gamma}, & \text{power-law decay} \\
        & B_{0}e^{-t/\tau} + B_{\rm m}, & \text{exponential decay}\\
        & 10^{12}\,{\rm G},& \text{constant}
    \end{aligned}
\right.
\end{equation}
\textcolor{black}{where $B_{0}=10^{14}~\rm{G}$} is the initial magnetic field, $B_{\rm m} = 5 \times 10^{11}$ G is the minimal magnetic field, $\gamma=1.6$, $\tau=\tau_{\rm d}/(B_{0}/10^{15}~\rm G)^{\gamma}$, $\tau_{\rm d}=10^{3}~\rm yr$   \citep{Dall’Osso2012}.
The solid, dashed, and dotted lines represent three different magnetic field evolution scenarios: exponential decay, power-law decay, and constant field, respectively.  The left panel shows the spin evolution tracks for ${\rm log}~t\, ({\rm yr}) > 7.25$. Before this time, the NS remains in a slow spin-down phase ($phase~a$).
The right panel depicts the magnetic field evolution over the lifetime of the companion star, with the red dashed box highlighting the time interval corresponding to the time range in the left panel. It can be seen that the early strong magnetic field has no significant impact on the final spin period of the NS. 

\textcolor{black}{There are considerable uncertainties in the birth properties of NSs, particularly regarding their spin periods and surface magnetic field strengths. Observations indicate a broad distribution of the initial spin periods, typically ranging from a few milliseconds to hundreds of milliseconds \citep{Faucher2006, Popov2012}. For NSs born with rapid rotation and strong magnetic fields, significant angular momentum can be carried away by pulsar winds during early evolutionary stages, substantially affecting the spin-down process. This angular momentum loss results from electromagnetic torques driven by relativistic outflows along open magnetic field lines \citep{Goldreich1969,Spitkovsky2006,Philippov2015}. However, as the open magnetic flux and particle outflow efficiency decline rapidly with spin-down, the influence of pulsar winds becomes negligible once the spin period exceeds $P \sim 1$ s \citep{Contopoulos1999}. Using the spin-down torque formula $N \propto (1 + \sin^2\alpha)$ \citep{Spitkovsky2006}, we estimate that - regardless of the specific values of the initial spin period or magnetic inclination angle ($\alpha$), a typical NS with magnetic field $B=10^{12}$ G will evolve to $P > 1$ s within $10^7$ yr. Various pulsar wind models have been proposed with different assumptions about spin-down mechanisms \citep[e.g.,][]{Contopoulos1999, Spitkovsky2006, Kalapotharakos2012, Parfrey2016}. However, these model-dependent differences become insignificant once the system enters the slow-rotator regime. The associated uncertainties primarily affect the early phase of spin evolution and have minimal impact on the long-term behavior that is the focus of this study.
}

\section{Conclusions} \label{sec:conclusion}
In this study, we explore the binary pathway for the formation of ULPPs. 
The possibility of ULPP formation in binaries is quantified by combining simulations with BPS and {\tt MESA} for binary evolution and NS spin evolution.
Our calculations show that the slow spin periods of these pulsars can be attributed to a two-stage process: (1) a prolonged spin-down phase induced by wind-fed accretion in wide HMXBs, followed by (2) the disruption of the binary due to the supernova explosion of the companion star. These binaries should have initial orbital periods longer than about $10^3$ days. Otherwise, Roche-lobe overflow onto the NS will significantly spin up the NS to short periods.

We estimate the formation rate of such NSs to be $\sim7.8\times 10^{-3}$\,$\rm yr^{-1}$, and $\sim 10^{-6}$\,$\rm yr^{-1}$ for those with $P_{\rm s}>1000$ s. In comparison, the total birthrate of the NS population in the Galaxy was inferred to be $\sim 10^{-2}$ yr$^{-1}$ \citep{Faucher2006,Lorimer2006,Keane2008}. It is difficult to estimate the total number of the potential ULPP population, not only because the small observational sample inhibits credible constraint on the birthrate of ULPPs, but also because it is still unclear whether ULPPs possess dipolar magnetic field configuration, what their radiation mechanism is, and how long their pulsar lifetimes are. 

According to \cite{Wang1982}, binary X-ray pulsars which have experienced extensive spin-down would have small oblique angles between the spin and magnetic axes, provided that the magnetic field is dipolar. In this situation one might expect that for long period NSs with the binary origin, one of their magnetic poles may remain hidden from view. In addition,
after the second supernova, both NSs are possibly located within the supernova remnant. Detection of an ULPP with a young NS in the same remnant could serve as evidence to test the binary origin scenario.

\begin{acknowledgments}
We are grateful to an anonymous referee for helpful comments. This work was supported by the National Key Research and Development Program of China (2021YFA0718500) and the Natural Science Foundation of China under grant Nos. 12041301 and 12121003.
\end{acknowledgments}

\software{BPS \citep{Hurley2000,Hurley2002}, MESA \citep{Paxton2011,Paxton2013,Paxton2015,Paxton2019}. 
}

\bibliography{ref}{}
\bibliographystyle{aasjournal}

\end{document}